\documentclass[twocolumn,amsmath,amssymb,aps,floatfix, longbibliography,superscriptaddress]{revtex4-1}
\usepackage{graphicx}
\usepackage{dcolumn}
\usepackage{bm}
\usepackage{amssymb}

\begin{document}


\title{Testing constitutive relations by running and walking on cornstarch and water suspensions}

\author{Shomeek Mukhopadhyay}
\affiliation{Department of Mechanical Engineering and Materials Science, Yale University, New Haven, CT 06520}
\author{Benjamin Allen}
\affiliation{Department of Mechanical Engineering and Materials Science, Yale University, New Haven, CT 06520}
\affiliation{School of Natural Sciences, University of California, Merced, California 95343}
\author{Eric Brown}
\affiliation{Department of Mechanical Engineering and Materials Science, Yale University, New Haven, CT 06520}
\affiliation{School of Natural Sciences, University of California, Merced, California 95343}

\date{\today}

\begin{abstract}
The ability of a person to run on the surface of a suspension of cornstarch and water has fascinated scientists and the public alike. However, the constitutive relation obtained from traditional steady-state rheology of cornstarch and water suspensions has failed to explain this behavior.  In a previous paper, we  presented an averaged constitutive relation for impact rheology consisting of an effective compressive modulus of a system-spanning dynamically jammed structure \cite{MMASB17}.  Here, we  show that this constitutive model can be used to quantitatively predict, for example, the trajectory and penetration depth of the foot of a person  walking or running on cornstarch and water.  The ability of the constitutive relation to predict the material behavior in a case with different forcing conditions and flow geometry than it was obtained from suggests that the constitutive relation could be applied more generally. We also present a detailed calculation of the added mass effect to show that while it may be able to explain some cases of people running or walking on the surface of cornstarch and water for pool depths $H >1.2$ m and foot impact velocities $V_I> 1.7$ m/s, it cannot explain observations of people walking or running on the surface of cornstarch and water for smaller $H$ or $V_I$.
\end{abstract}

\maketitle
\label{sec:Intro}
 
Discontinuous Shear Thickening (DST) suspensions  exhibit a remarkable effect in which they behave like typical liquids at low shear rates, but when sheared faster, resistance to flow can increase discontinuously with shear rate \cite{Ba89,BJ14}.  DST suspensions can also exhibit solid-like properties such as cracking \cite{RMJKS13}. DST has been observed in a large variety of concentrated suspensions of hard, non-attractive particles, and is inferred to be a general feature of such suspensions \cite{Ba89, BJ12, BJ14,BFOZMBDJ10}. DST suspensions also support large stresses under impact, one example of which is the ability of a person to walk or run on the surface of a pool filled with a suspension of cornstarch and water  \cite{youtube_running, BJ14}.  The impact response of such fluids is of practical interest for impact protection gear because of their strong response during impact while remaining fluid and flexible otherwise \cite{LWW03,D3O}. The purpose of this paper is to test a constitutive relation that relates the force on an impactor to its displacement into a DST suspension \cite{MMASB17}.  We demonstrate its generality by using it to quantitatively describe the ability of people to run or walk on the surface of cornstarch and water. Such a constitutive relation may aid in the development of materials for impact protection applications.
  

The impact response of DST suspensions has been long assumed by the scientific community to be a direct consequence of shear thickening based on steady-state rheology \cite{WB09}. 
In qualitative support of this argument, observations without the benefit of controlled laboratory conditions do show that it is possible for a person to run on the surface of cornstarch and water suspensions as it were a solid.  However,  at a lower foot impact velocity the suspension may remain more liquid-like so that the person sinks in \cite{BJ14,youtube_running} (See Supplementary Video 1). The rate-dependence of this liquid-like to solid-like transition is qualitatively similar to DST.  Observations also show that this effect happens only at high packing fractions, also  similar to DST  \cite{CPJCPL13}. However, it has never been quantitatively tested whether these observations are related to or can be explained by steady state rheological models of shear thickening. 
 
In steady-state rheology, shear thickening is defined by a range with a positive slope in the viscosity function $\eta(\dot\gamma) \equiv \tau/\dot\gamma$ as a function of shear rate, where $\tau$ is the shear stress and $\dot\gamma$ the shear rate in a steady-state shear flow.  The intent of such a constitutive relation is to predict flows with different forcing conditions, boundary conditions, and geometries.  The constitutive relation obtained from steady-state measurements indicates that suspensions of cornstarch and water can support steady shear and normal stresses up to $\sim 10^3$ Pa in a shear rate range where they are shear thickening, i.e.~before they become shear thinning (a negative slope of $\eta(\dot\gamma)$) at higher shear rates \cite{BJ12}. 
If we try to apply this  result from steady-state rheology to a person running on cornstarch and water, the predicted stress of $\sim10^3$ Pa is much less than needed for a person to be supported on the surface of the fluid,  based on a simple estimate of a person's weight distributed over the surface area of a foot ($\approx 4\cdot 10^4$ Pa).  Thus, the constitutive relation obtained from steady-state rheometer experiments fails to explain the strong response to impact.  It remains an open question whether our understanding of steady state DST can be extended to explain the strong impact response.


Recently an `added mass' model has been developed for impact response of dense suspensions, in which a `dynamically jammed' region forms ahead of the impactor in the fluid.  In this localized region, the suspension moves along with the impactor like a plug \cite{WJ12}.   The dynamically jammed region grows during the impact with a front which propagates away from the impactor \cite{WJ12,PJ14,HPJ16}.   There is a sharp velocity gradient at the front, which separates the dynamically jammed region from the surrounding fluid \cite{PJ14}. In a two-dimensional dry granular experiment the front velocity and width diverge at the same critical packing fraction as the viscosity curve of DST suspensions  \cite{WRVJ13}.  

In the  model for the added mass effect, the impact response of the suspension comes from an increasing suspension mass (i.e.~the `added mass') in the dynamically jammed region which moves with the impactor \cite{WJ12}.  This increasing mass slows down a free-falling impactor due to conservation of momentum.     
This model has been confirmed to quantitatively describe  the impact response of some high-speed projectiles into suspensions \cite{WJ12}.  However, to  significantly slow the impacting object by momentum conservation alone requires  large masses of fluid compared to the impacting object (or similarly, large depths of the fluid compared to the object's height).    The added mass model has not been quantitatively applied to other impact response problems.  The regime of thin fluid layers where the added mass effect is weak is also particularly important for the related problem of impact protection applications where thin layers of protective material are desired \cite{DHNWW07,LWW03}.


When the dynamically jammed region reaches the boundary, the stress on the impactor increases beyond the added mass effect \cite{PJ14, MMASB17}.  We found that  the stress increase follows immediately after particle displacement is observed at the boundary opposite impact, implying that the stress increase comes from deformation of the dynamically jammed region  \cite{ASMMB17}. We also observe dilation at this boundary in the same region where we find particle motion  \cite{ASMMB17}.  This observation is reminiscent of soils or dense granular materials and suggests a force transmission between particles along frictional contacts, as shear of a dense packing induces dilation as a result of particles pushing into and around each other.  This suggests the dynamically jammed structure could support a normal load  that is transmitted via frictional interactions across the system  when the dynamically jammed region spans from the impactor to a solid stationary boundary.  The system-spanning dynamically jammed structure was found to resist the impact with an effective compressive modulus $E$, which was measured as a function of weight fraction $\phi$, impact velocity $V_I$, and fluid height $H$ \cite{MMASB17}.  We now want to test whether this constitutive relation can be used to quantitatively describe stresses, deformations, and flows of the material with different forcing conditions, boundary conditions, and flow geometries than the experiments used to obtain the constitutive relation.  

In this paper, we use a person running and walking on the surface of cornstarch and water as a test case for the constitutive relation.  This is a good test of the generality of the constitutive relation because the flow conditions differ significantly from the ideal laboratory experiments.  Specifically, the impactor shape is different, and it has a varying velocity profile, which is initially a free-fall, followed by a slowing due to impact.  We compare to predictions of both the constitutive relation for the system-spanning dynamically jammed region, as well as the added mass model.

The remainder of the paper is organized as follows.    In Sec.~\ref{sec:methods} we explain the experimental methods used.  In Sec.~\ref{sec:running}  we show measured trajectories of a foot of a person walking, jogging, and running on cornstarch and water that can be compared to models.  In Sec.~\ref{sec:constitutive} we test the constitutive model obtained in our previous paper \cite{MMASB17} by showing it can quantitatively explain the ability of people to walk and run on the surface of cornstarch and water.    In Sec.~\ref{sec:addedmass}, we give a detailed  calculation of the added mass effect, and show that it cannot explain most observations of people walking or running on the surface of cornstarch and water, although it can dominate the impact responses at higher impact velocities.

\section{Materials and methods}
\label{sec:methods}

We used high-speed video to track the foot of a person  walking, jogging and running on a  suspension of cornstarch and water in an inflatable outdoor pool of length $1.8$ m, width $0.9$ m, and depth $H=100\pm 12$ mm, where the uncertainty in $H$ is due to the unleveled ground.  The person had mass $m=85$ kg, and a cross-sectional area on the bottom of one foot of $A=0.020$ m$^2$.  The cornstarch was purchased from Carolina Biological Supply, and mixed with tap water in a concrete mixer.  We inferred an effective weight fraction $\phi$ by placing a portion of the same suspension on a rheometer, and measured the critical shear rate $\dot\gamma_c=6\pm 2$ s$^{-1}$ at the onset of shear thickening.  We obtained $\phi$  via a conversion function $\phi(\dot\gamma_c)$ we fit for our lab conditions \cite{MB17}, which correspond to the lab conditions where the constitutive relation was obtained  \cite{MMASB17}.  We find $\phi=0.577\pm0.007$ on this scale, where the liquid-solid transition is at $\phi = 0.61$ \cite{MB17}.  This weight fraction is in the range of strong shear thickening \cite{BJ09}, and in the range where the effective compressive modulus $E$ reaches a plateau independent of weight fraction \cite{MMASB17}.  The density of the suspension at the weight fraction is $\rho=1200\pm20$ kg/m$^3$ \cite{MMASB17}.

\section{Experimental results}
\label{sec:running}

\begin{figure}  
\centering
{\includegraphics[width=0.475 \textwidth]{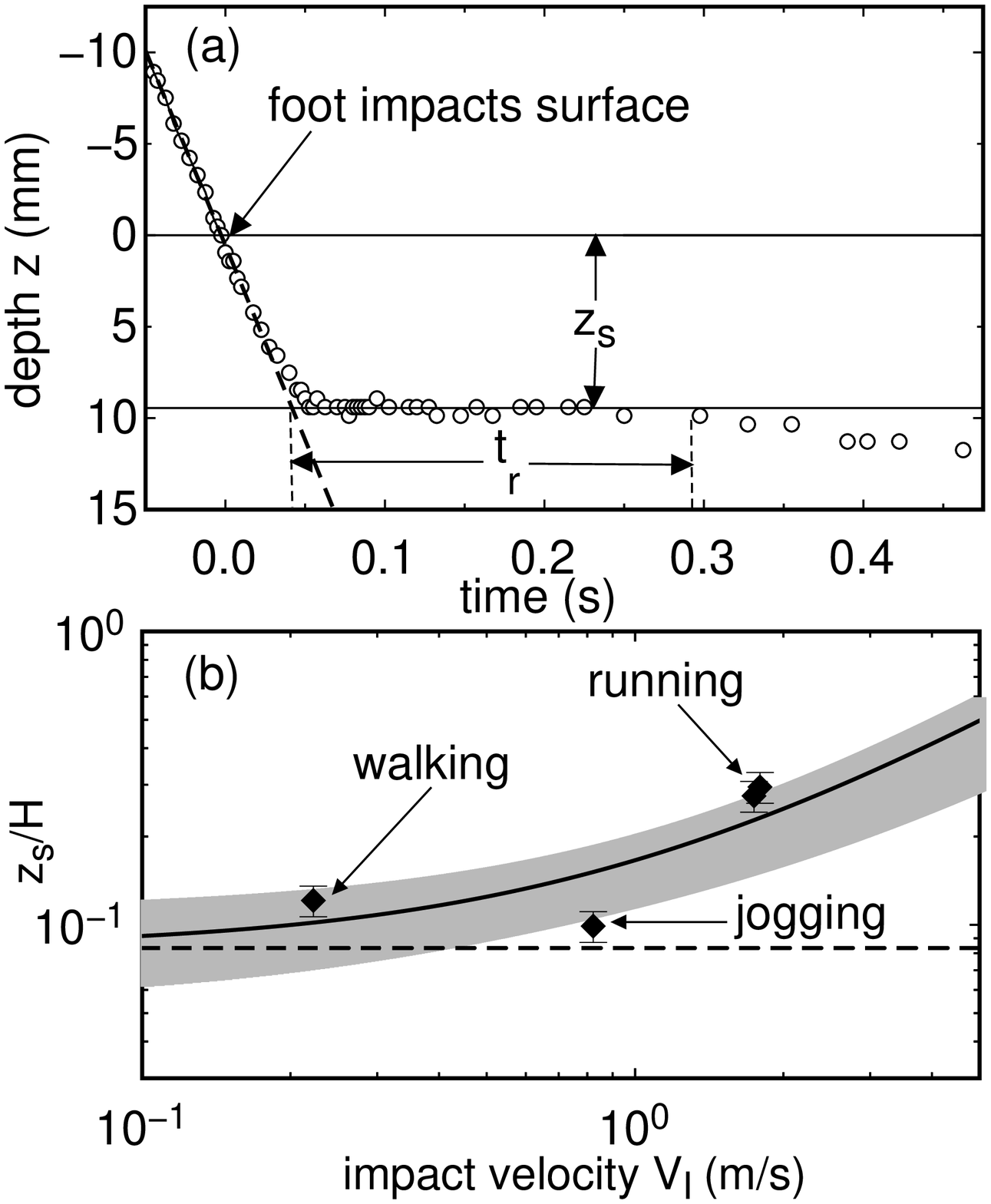}}
\caption {(a) Experimental observation of the penetration depth $z$ as a function of time for a foot  of a person walking on cornstarch and water with a foot impact velocity $V_I = 0.2$ m/s. The foot came to an abrupt stop below the surface ($z=0$) at depth $z_s$.  This delay in force response is similar to the delay in controlled velocity experiments \cite{MMASB17}. (b)  Symbols: Measurements of the stopping distance $z_s$ as a function of $V_I$. Dashed line: model prediction for the penetration depth $z_F$ when the front  of the dynamically jammed region reaches the boundary. Solid line: model prediction for the total stopping distance $z_s = z_F+\Delta z$, where $\Delta z$ is the additional displacement due to the kinetic energy of the impact deforming the system-spanning dynamically jammed region.   Gray band: error on the solid line prediction. The agreement between the measurements and the model prediction shows that walking, jogging and running on cornstarch and water  can result from the system-spanning  dynamically jammed region.
} 
\label{fig:stoppingdistance}
\end{figure}

We tracked the  vertical displacement $z$ of a person's foot.   This was done by following a point on the foot that moved with the point of deepest penetration with a resolution of $ 0.2$ mm.  The penetration depth of the foot as a function of time is shown in Fig.~\ref{fig:stoppingdistance}a for a person walking on cornstarch and water.  A striking observation is that the velocity of the foot did not change as it first penetrated the surface. Thus there was no significant stress acting on the foot until it came to an abrupt stop at depth $z_s \approx 10$ mm below the surface.  This is qualitatively consistent with a delay before a sharp stress increase due to the time it takes for the dynamically jammed region to span between solid boundaries \cite{MMASB17}.  On the other hand, it is qualitatively inconsistent with models which predict a smooth deceleration, such as a bulk viscosity or an added mass effect \cite{WB09, WC14, WJ12}.  We will revisit predictions of the added mass effect in more detail in Sec.~\ref{sec:addedmass}.

The late-time trajectory of the foot shown  in Fig.~\ref{fig:stoppingdistance}a  reveals information about relaxation of the suspension.  After the foot stopped, it remained stationary for an additional time $t_r=0.25$ s,  before it began to slowly sink again.   This  relaxation corresponds  to the common observation that the dynamically jammed state melts  from a solid-like state back into a liquid-like state some time after impact. This  timescale is larger than the range of stress relaxation times measured in analogous rheometer experiments (0.013-0.08 s) on cornstarch and water at weight fractions in the range $\phi=0.577\pm0.007$.  This relaxation time corresponds to the characteristic timescale of an exponential decay, so at $t_r=0.25$ s, the stress would have time to decay down to a fraction of its peak value during impact. This timescale $t_r = 0.25$ s is also longer than a typical runner's step duration of 0.15 s  \cite{WSBW00},  but shorter than  the walker's step duration in Fig.~\ref{fig:stoppingdistance}a,  at least at this weight fraction.  This observation could explain why if a person stands on the  surface of a suspension too long  before taking another step,  they sink  into the suspension.  To make quantitative predictions of this threshold, the relaxation time would have to be measured more precisely, and the sinking behavior would have to be modeled, which are beyond the scope of this paper.     
 
The main feature in Fig.~\ref{fig:stoppingdistance}a that can be compared to a quantitative model prediction for impact response is the stopping distance $z_s$.  While the foot  was angled by up to 15 degrees when it impacted the surface at $z=0$, the final  orientation of the foot  at the penetration depth $z_s$ was parallel with the suspension surface to within our measurement resolution of $0.3$ degrees.  This results in a variation in $z_s$ of less than $0.5\%$  when comparing to  impact experiments with parallel surfaces, which is relevant if  propagation  of the dynamically jammed region depends mainly on the point of first contact.  
We obtain the impact velocity $V_I$  from the measurements in Fig.~\ref{fig:stoppingdistance}a by fitting a straight line to $z(t)$  before impact.   The stopping distance $z_s$ is obtained from a fit to the plateau in $z$ as shown in Fig.~\ref{fig:stoppingdistance}a.    We plot  our  measured $z_s/H$ for a person's foot running, jogging and walking on cornstarch and water as a function of $V_I$ in Fig.~\ref{fig:stoppingdistance}b, where the dominant error is from the uncertainty on $H$.

 \section{Testing the constitutive relation for the system-spanning dynamically jammed region}
  \label{sec:constitutive}
  
In this section we test the averaged constitutive rheology that was obtained from controlled velocity experiments \cite{MMASB17} by applying it to the results of Sec.~\ref{sec:running}.  

The constitutive relation consists of an effective compressive modulus $E=(H-z_F)d\tau/dz$, but only after the dynamically jammed region spans between boundaries for $z>z_F$\cite{MMASB17}.   For $z<z_F$, there is no stress on the impactor due to the deformation of the dynamically jammed region, although there may be much smaller background effects due to buoyancy and other forces that will be ignored here \cite{MMASB17}.  The modulus $E$ has unusual parameter dependences, for example depending on the fluid height $H$.  The stiffness per unit area $d\tau/dz$ also depends on impact velocity and weight fraction, and reaches a plateau in the limit of large impact velocities $V_I$ and large weight fractions $\phi$, where  $d\tau/dz=64\pm 9$ MPa/m for $V_I \ge 100$ mm/s and $\phi\ge0.57$ \cite{MMASB17}. This plateau includes the parameter range of our experiments.  The front of the dynamically jammed region moves at velocity $V_F = kV_I$, where $k$ is another material parameter. The delay depth $z_F$ corresponds to a depth of the impactor at the time that the dynamically jammed region first spans to the bottom of the impactor, and can be expressed as $z_F = H/k$ for constant $V_I$.  The value of $k$ also depends on $V_I$ and $\phi$, and reaches a plateau with constant $k=12\pm 4$ for $V_I \ge 100$ mm/s and $\phi\ge0.57$ \cite{MMASB17}.   This plateau again includes the parameter range of our experiments.   While these parameters were measured as a function of weight fraction up to the jamming transition, impact velocities were only measured up to 0.4 m/s \cite{MMASB17}.  For the purposes of testing a model, we extrapolate by using the same plateau values of $d\tau/dz$ and $k$ for higher velocities.  This extrapolation can be partially validated for $k$, where it was found to be independent of $V_I$ for higher velocities as well \cite{WJ12}.

We now use the averaged constitutive relation from the controlled velocity experiments to build a quantitative model  that can predict the stopping distance $z_s$ of an  object impacting  the surface of a cornstarch and water suspension. The majority of the stress response is expected  due to the deformation of the dynamically jammed region when it spans to the bottom boundary after $z=z_F$. The inertia of the impactor will lead to a deformation $\Delta z$ of the dynamically jammed region, determined by equating the kinetic energy of the impactor with the work done to deform the dynamically jammed region of compressive modulus $E$ and initial height $H-z_F$ (i.e.~when $z=z_F$).  To obtain a relative simple, lowest-order calculation, we evaluate this energy balance in the limit of small deformation 

\begin{equation}
\frac {m V_{I}^{2}}{2}=\frac {EA(\Delta z)^{2}} {2(H-z_F)},
\label{eqn:energybalance}
\end{equation}

\noindent where $A$ is the cross-sectional area of the impacting object, and $m$ is the mass of the impacting object. The total stopping distance is predicted to be $z_s= z_F + \Delta z$, including both the penetration depth $z_F$ when the dynamically jammed region spans to the boundary, and the further  penetration $\Delta z$ equal to the deformation of this structure afterward.   Using this expression for $z_s$,   Eq.~\ref{eqn:energybalance}, and the relations $z_F = H/k$ and $E=(H-z_F)d\tau/dz$ yields

\begin{equation} 
\frac{z_s} {H}= \frac {1} {k} + \frac{V_I}{H} \left [\frac{m} {A(d\tau/dz)} \right ]^{1/2} \ .
\label{eqn:zs}
\end{equation}

  We compare this prediction with the observations of a person's foot in Fig.~\ref{fig:stoppingdistance}b, where the model parameter values are chosen to match the  experiment: $H=0.10 \pm 0.012$  m,  a person of mass $m=85$ kg, and foot cross-sectional area $A=0.020$ m$^2$.  We use the values $d\tau/dz=64\pm 9$ MPa/m and $k=12\pm 4$ for the large $\phi$ and $V_I$ limit as mentioned above \cite{MMASB17}.    To confirm that the model covers the stress range of the experiment, we checked that the fit range of the modulus $E$  \cite{MMASB17} included up to the maximum stress expected on the foot.  The  maximum stress expected on the foot  is calculated from Eq.~\ref{eqn:energybalance} as $\tau = E\Delta z/(H-z_F) = mV_I^2/A\Delta z = 8\cdot10^5$ Pa, which is within the range of the fit of $E$ \cite{MMASB17} (for this calculation the maximum $V_I$ and $\Delta z$ were obtained from Fig.~\ref{fig:stoppingdistance}b). 
  
  The dominant errors in the model parameters are the run-to-run variations in the measurements of $E$ ($14 \%$)  and $k$ ($36 \%$) corresponding to the standard deviation of measurements.  An assumption in Eq.~\ref{eqn:energybalance} is that  our control parameter $V_I$  is a good approximation for the center of mass velocity $V_0$ (the latter would be appropriate in Eq.~\ref{eqn:energybalance}).  We took videos of people running and found that $V_I$ tends to overestimate $V_0$  by 20\% to 50\%,  which  is represented with a negative error bar on the predicted $\Delta z$ of 50\%  if we take these percentages of typical numbers for all $V_I$.  In the model we assume that the area $A$ stays constant, however during our measurement as much as $45 \%$ of the footprint was still outside of the suspension  when the impactor reached $z_F$,   which is represented by a positive error of 22\% on the predicted $\Delta z$.    After  all of these errors are included in the error bars shown in Fig.~\ref{fig:stoppingdistance}b, the predicted $z_s/H$ is statistically consistent with the measurements within an average of about one standard deviation.  The root-mean-square difference between the prediction and data is 26\%, within the run-to-run standard deviation of 30\% in stress. This agreement quantitatively confirms --  albeit with a large error bar -- that the penetration depth $z_s$ of  a person's foot while walking, jogging, or running across the surface  of cornstarch and water can be predicted based on the constant-velocity impact experiments \cite{MMASB17}

Equation \ref{eqn:zs} has two distinct scaling regimes.  For small $V_I$,  very little kinetic energy has to be absorbed, so $z_s$ is dominated by the delay depth $z_F$ required for the dynamically jammed region to span between solid boundaries (the first term in Eq.~\ref{eqn:zs}, plotted as the dashed line in Fig.~\ref{fig:stoppingdistance}b). In this regime $z_s \approx H/k$,  and the values shown in Fig.~\ref{fig:stoppingdistance}b in this limit are quite general over the wide parameter range in $\phi$, $V_I$, $H$, and $d$ in which  $k$ and $d\tau/dz$ are  constant \cite{MMASB17}.  For large $V_I$, the kinetic energy has  to be absorbed by compression of the dynamically jammed structure, so $z_s$ is dominated by $\Delta z$ (the second term in Eq.~\ref{eqn:zs}). In this regime, $z_s$ is independent of $H$.  Considering  that the factor $(m/A)^{1/2}$  doesn't change much from person to person, and $d\tau/dz$ is constant in the large-$\phi$ plateau, we can approximate $z_s \approx 8\times10^{-3} V_I$ s over a wide parameter range for large $V_I$.

 The model of Eq.~\ref{eqn:zs} assumes the impact is at high enough velocity that the contribution of the change in gravitational potential energy $mg\Delta z$ over the course of the impact makes a negligible contribution to $\Delta z$.  It could easily be included as an additional term on the left side of Eq.~\ref{eqn:energybalance}.  This term would become dominant over the kinetic energy term at small $V_I$.  In our parameter range, the correction to $\Delta z/H$ from this term is only 1\% in the limit of $V_I =0$, so is small compared to the dominant term $z_F$ and safe to ignore here

\section{The added mass effect in impact response}
\label{sec:addedmass}

A competing model for explaining the ability of people to run on the surface of cornstarch and water is the added mass effect. While it is not needed to explain the results in Fig.~\ref{fig:stoppingdistance}, we expect it to be a dominant effect in different parameter regimes.  In this section we perform detailed calculations of the added mass effect using the model of Waitukatius \& Jaeger \cite{WJ12}  to determine under what conditions it can explain the ability of people to run on cornstarch and water.  We start with general trajectories for rigid free falling objects in Sec.~\ref{sec:addedmass_derivation}.  We develop a lower bound on the  required fluid height $H_c$  for a person to run on the surface of cornstarch and water in Sec.~\ref{sec:bounds}.  In Sec.~\ref{sec:bounds_trajectories}, we combine the bounds with the trajectories to develop more precise bounds on the  required fluid height $H_c$, and a minimum foot impact velocity $V_c$  for a person to run on the surface of cornstarch and water.

\subsection{Added mass effect for rigid free-falling objects}
\label{sec:addedmass_derivation}

The growth of the added mass over time slows down an impacting object due to conservation of momentum with a variable mass \cite{WJ12}.  To characterize the trajectories of rigid free-falling objects, we start with the time-dependent  force balance equation with conservation of momentum for a variable mass
 
 \begin{equation} 
 m_0 g = [m_0+m_a(t)] \frac{dV_a}{dt} + V_a(t)\frac{dm_a}{dt}  
\label{eqn:forcebalance}
\end{equation}

\noindent where $m_0$ is the mass of the impacting object, $V_a$ is the time-dependent velocity of the added mass (we assign downward velocity to be positive), and $m_a$ is the time-dependent added mass that moves with the impacting object.   We use $V_a$  to represent both the velocity of the added mass as well as the impacting object, as previous experiments showed that them to be the same while they are in contact \cite{PJ14}.  This corresponds to an  uncompressed dynamically jammed region  before it spans between solid boundaries.  We assume for ease of calculation that the impacting object is rigid, so the center of mass velocity is the same as its impacting surface, and thus $V_a$.   This  assumption will not hold, for example,  for a person running on the surface  of cornstarch and water,  as the center of mass velocity can differ from the foot velocity due to bending of the leg. We  will calculate bounds that  account for this lack of rigidity in Sec.~\ref{sec:bounds}.  As in Waitukaitus \& Jaeger \cite{WJ12}, we ignore effects of viscous drag which may increase with the added mass.  We also ignore effects of buoyancy on the added mass since  the density difference between the liquid and dynamically jammed regions is small \cite{HPJ16}.   Rearranging Eq.~\ref{eqn:forcebalance} to isolate $dV_a/dt$, integrating over time, and using the initial condition $V_a(t=0) = V_I$ at the time of impact yields
 
   \begin{equation} 
 V_a =V_I - \int_0^{t} \left[\frac{V_a}{m_0+m_a}\frac{dm_a}{dt}   -  \frac{m_0 g}{m_0+m_a} \right]dt \ .
\label{eqn:equationofmotion}
\end{equation}
 
The geometry of the added mass $m_a$ was empirically fit by a frustrum shape based on the force response on free-falling objects \cite{WJ12}.   Thus, the added mass can be written as a function of penetration depth $z$ as
  
  \begin{equation} 
m_a= \frac{0.37\pi \rho }{3} \left(\frac{D}{2}+kz \right)^2 kz \ ,
\label{eqn:addedmass} 
\end{equation}

 \noindent where $\rho$ is the fluid density, $D$ is the impactor diameter, and $k$ represents the ratio of front velocity $V_F$ to impact velocity $V_I$.  In practice, $k$  is a free parameter  which depends on weight fraction $\phi$.  $dm_a/dt$ is obtained from the analytic derivative of Eq.~\ref{eqn:addedmass}, making use of the identity $V_a = dz/dt$
  
  \begin{equation} 
\frac{dm_a}{dt} = 0.37 \pi\rho kV_a \left(k^2z^2 + \frac{2kzD}{3} + \frac{D^2}{12}\right) \ .
\label{eqn:dma_dt} 
\end{equation}

To  reduce the number of  parameters in Eqs.~\ref{eqn:equationofmotion}-\ref{eqn:dma_dt}, we introduce a characteristic length scale $z_0\equiv (3m_0/0.37\pi\rho)^{1/3}$.  This is a characteristic depth of the dynamically jammed region (corresponding to $z_0=kz)$, specifically when $m_a=m_0$ in Eq.~\ref{eqn:addedmass} in the limit of small $D\rightarrow0$.   With this scale, we define the non-dimensional velocity  $\tilde V_a = V_a/V_I$, added mass $\tilde m_a = m_a/m_0$, impactor diameter $\tilde D = D/z_0$, impactor depth $\tilde z = zk/z_0$, time $\tilde t = tV_I/kz_0$, and gravity $\tilde g = g kz_0/V_I^2$.  The different normalizations for $\tilde D$ and $\tilde z$ were chosen to allow for a minimum number of control parameters in dimensionless form, while the main material parameter $k$ is hidden in the  normalizations of $\tilde t$ and $\tilde z$.  Using this nondimensionalization, and inserting Eq.~\ref{eqn:dma_dt} into \ref{eqn:equationofmotion} yields the dimensionless form of Eq.~\ref{eqn:equationofmotion}

\begin{equation} 
\tilde V_a =1 -\int_0^{\tilde t}  \left[\frac{3\tilde V_a^2 \tilde z^2}{1+\tilde m_a}\left(1+\frac{2\tilde D}{3\tilde z} +\frac{\tilde D^2}{12\tilde z^2}\right)  - \frac{\tilde g}{1+\tilde m_a}\right] d\tilde t  
\label{eqn:nondimensional}
\end{equation}
 
 \noindent and the dimensionless form of Eq.~\ref{eqn:addedmass}
 
   \begin{equation} 
\tilde m_a= \left(\frac{\tilde D}{2}+\tilde z \right)^2 \tilde z \ . 
\label{eqn:addedmass_dimensionless} 
\end{equation}
 
 \noindent These equations reduce the system down to two dimensionless parameters: $\tilde D$ and $\tilde g$.  
 
 With the non-dimensionalization of Eqs.~\ref{eqn:nondimensional} and \ref{eqn:addedmass_dimensionless}, a set of general trajectories may be expressed in terms of $\tilde V_a(\tilde t)$.  This is obtained by a  numerical integration of the implicit Eq.~\ref{eqn:nondimensional} using Eq.~\ref{eqn:addedmass_dimensionless} for $\tilde m_a$, while simultaneously integrating $\tilde z = \int \tilde V_a d\tilde t$ (since $z \equiv \int V_a dt$).    These trajectories essentially reproduce what was calculated by Waitukaitus \& Jaeger \cite{WJ12}, but we additionally note that in this dimensionless form this and other curves are universal  for rigid impacting objects.

 \subsubsection{Solutions in the limit $\tilde D=0$, $\tilde g=0$}

 \begin{figure}  
\centering
{\includegraphics[width=0.475 \textwidth]{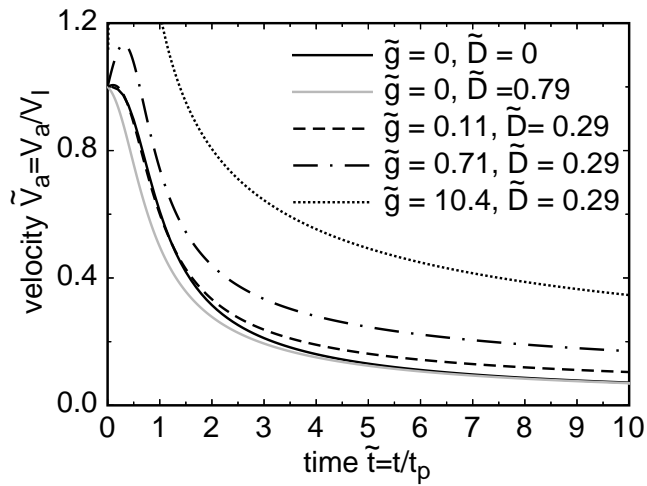}}
\caption {
Dimensionless velocity trajectories for free-falling rigid objects impacting into a suspension, where the objects are slowed due to the added mass effect. Trajectories are shown for different values of the dimensionless gravity $\tilde g$ and impactor diameter $\tilde D$.  Solid black line: $\tilde g = 0$, $\tilde D=0$, corresponding to the limits of large impact velocity $V_I$ and dense/thin impactors. Solid gray line: $\tilde g = 0$, $\tilde D=0.79$, corresponding to an aspect ratio 1 cylinder at large $V_I$.  Disconnected lines: $\tilde D=0.29$,  corresponding to a person running at $V_I=2$ m/s (dashed line, $\tilde g = 0.11$), jogging at $V_I=0.8$ m/s (dashed-dotted line, $\tilde g = 0.71$), or walking at $V_I=0.2$ m/s (dotted line, $\tilde g = 10.4$) on cornstarch and water. 
} 
\label{fig:integration}
\end{figure}

  A result for $\tilde V_a(\tilde t)$ in the limit of $\tilde g=0$ and $\tilde D=0$ is shown in Fig.~\ref{fig:integration}  as the solid black curve.    A few notable scales can be identified from the trajectories of $\tilde V_a(\tilde t)$ in this limit. The added mass initially grows rapidly over time, but does not significantly decelerate the impacting object until $m_a$ becomes comparable to $m_0$.  The velocity is reduced by half when $m_a = m_0$ at impactor depth $z_0/k$ (corresponding to a front depth $z_0$).   There is also an inflection point in $\tilde V_a(\tilde t)$ where deceleration is maximum at $t = 0.71z_0/kV_I$.    Thus, the timescale $t_p \equiv z_0/kV_I$ and length scale $z_p \equiv z_0/k$  are rough scales  over which most of the slowdown from the added mass effect occurs.   Beyond these scales, the combined object has significantly more mass ($m_0+m_a$) than the impactor itself ($m_0$), and thus is harder to decelerate.  

 The added mass effect is a variation on inertial mass displacement in impact response, in which in which the force in response to impact  scales as $F_p \sim m_0 V_I^2/z_0$.  This scaling is found for a large variety of fluids, ranging from liquids, to suspensions, to dry granular beds \cite{Schlicting60}.    For the added mass effect, we find the peak force on the impactor is $F_p = max(m_0 dV_a/dt) = 0.61k m_0 V_I^2/z_0$ in the limit of $\tilde g=0$ and $\tilde D=0$.    This is larger by about a factor of $k$  than for most other fluids, which is what makes the added mass effect stronger than a typical inertial impact response.

 \subsubsection{Perturbations on the limiting solution}
 
The dimensionless parameter  $\tilde D$ is an aspect ratio representing the effective diameter of the impactor divided by the depth of  front of the dynamically jammed region required for $m_a=m_0$.  $\tilde D$ is small for impacting objects which are much denser than the fluid and/or tall and thin.   As an example for comparison, $\tilde D=0.79$ for a cylinder of aspect ratio 1 and the same density as the fluid.  A trajectory for this case is shown as the  solid gray curve in Fig.~\ref{fig:integration}.  For $\tilde D \ge 0$, the impacting object initially decelerates faster due to the larger initial cross-sectional area of the added mass, but as $m_a \stackrel{>}{_\sim} m_0$ ($\tilde V_a \stackrel{<}{_\sim} 0.5$) the initial contact area becomes less relevant compared to the total added mass, and the trajectory approaches the $\tilde D =0$ trajectory.

The relative contribution of gravity is  characterized by $\tilde g$,  which becomes smaller mainly as impactor velocity $V_I$ increases, so  $\tilde g =0$ can be taken as the limit where $V_I$ is large.  To provide some examples, we use parameters corresponding to a person, although these trajectories would only apply to a single step and still ignore deformation of the body during impact.  We obtain an effective diameter for  the foot by assuming the same contact area $A = 0.020$ m$^2$ as the foot used in Fig.~\ref{fig:stoppingdistance} and defining the effective foot diameter $D = 2\sqrt{A/\pi} = 0.16$ m.  

For a person of mass $m_0 =85$ kg, suspension density $\rho=1200$ kg/m$^3$, and effective foot diameter $D=0.16$ m, we obtain $z_0 = 0.56$ m and $\tilde D = 0.29$. Since $\tilde D \propto A^{1/2}(\rho/m_0)^{1/3}$ does not vary much from person to person, and the trajectory of $V_a$ does not change much with $\tilde D$,  this value $\tilde D=0.29$ is used for all further calculations of the added mass effect on a person.  We obtain values of $\tilde g$ for a person walking, 

Free-fall trajectories are shown corresponding to the parameters of a person walking, jogging, and running on cornstarch and water are shown in Fig.~\ref{fig:integration}.  We use $\tilde D=0.29$ and different values of $\tilde g$ for $k=12$ \cite{WJ12, MMASB17} and impact velocities for running ($\tilde g = 0.11$, $V_I=2$ m/s), jogging ($\tilde g =0.71$, $V_I=0.8$ m/s), and walking ($\tilde g=10.4$, $V_I=0.2$ m/s), as measured in Fig.~\ref{fig:stoppingdistance}.   For small $V_I$ ($\tilde g \stackrel{>}{_\sim} 1$), including the jogging and walking steps, the person is expected to continue accelerating downward on a nearly free-fall trajectory initially until the added mass builds up enough to be comparable to $m_0$. 
For most of the examples shown in Fig.~\ref{fig:integration}, specifically for $\tilde g < 1$ (large $V_I$) and $\tilde D < 1$, the trajectories are similar to  the $\tilde D=0$, $\tilde g=0$ limit.  This includes impact velocities corresponding to jogging speeds and faster, and impactor aspect ratios of 1 and thinner.  In these cases, it is fair to consider the $\tilde D=0$, $\tilde g=0$ trajectory as a limiting case that other trajectories converge to,  while $\tilde D$ and $\tilde g$ are small perturbation parameters.

\subsection{A lower bound on the fluid height for running on cornstarch and water due to the added mass effect}
\label{sec:bounds}

The trajectories in Fig.~\ref{fig:integration} assume the added mass can grow indefinitely.  The added mass contribution could be much weaker if the growth of the added mass is stopped by reaching a solid boundary, i.e.~when $kz=H$.   At this point, the added mass contribution would vanish, and the system-spanning dynamically jammed region would likely become dominant, as observed in Fig.~\ref{fig:stoppingdistance}.  Here, we obtain  a lower bound  on the  minimum height $H_c$ of the fluid required for a person to run on the surface of cornstarch and water due to the added mass effect.

 First, we calculate an expression for the available added mass $m_c$ in a pool of height $H_c$.    Plugging in the maximum extent of the added mass, $kz = H_c$ into Eq.~\ref{eqn:addedmass} yields the available added mass
 
 \begin{equation} 
 m_c= \frac{0.37\pi\rho }{3}\left(\frac{D}{2}+H_c\right)^2H_c \ .
\label{eqn:addedmass_bound}
\end{equation}

The minimum fluid height $H_c$ could be calculated  implicitly for a given $m_c$ using Eq.~\ref{eqn:addedmass_bound}. To do so, we need an expression for the added mass $m_c$ required to support a person  of mass $m_0$ to equate to Eq.~\ref{eqn:addedmass_bound}.  While Eq.~\ref{eqn:nondimensional} gives a general solution for impact response of a rigid free-falling object from the added mass effect, it does not account for internal motion of impacting objects.  In particular, for a person to run on the surface of cornstarch and water requires  actively pushing off of the surface with one leg before pushing down with another to get a force from the added mass effect.  The decoupling of the foot velocity from the center-of-mass velocity is necessary for it to be possible to  obtain an upward motion.  Otherwise, the person would always sink, and persistent running on the surface of cornstarch and water (in practice, until the person gets tired) would be impossible.  An expression for the required added mass $m_c$ can be calculated from momentum conservation, balancing the change in momentum of the person at the times of foot impact and liftoff from the surface  with the momentum of the added mass and the impulse from gravity: 

\begin{equation} 
V_0 m_0=-m_0\sqrt{V_0^2+2gz_{\Delta t}} + V_{\Delta t} m_c - m_0 g\Delta t \ .
\label{eqn:momentum}
\end{equation}

\noindent $V_0$ is the center-of-mass velocity of the person at the time of foot impact.     The terms on the right-hand side represent respectively the momentum of the person  at the time the foot lifts off the surface a time $\Delta t$ later,  the momentum of the  added mass with velocity $V_{\Delta t}$ at the time of liftoff, and the impulse from gravity on the person over the time $\Delta t$.  The liftoff velocity $\sqrt{V_0^2+2gz_{\Delta t}}$ on the right-hand side is determined by ballistic motion between steps.   Liftoff occurs at a depth $z_{\Delta t}$ lower than the quiescent fluid surface, resulting in a potential energy difference that has to be made up in the difference between liftoff and impact velocities.  Persistent running  on the surface requires each step to have the same $V_0$ and so must satisfy Eq.~\ref{eqn:momentum}.

To obtain an explicit expression for the required added mass,   we algebraically rearranging Eq.~\ref{eqn:momentum} as

\begin{equation} 
 \frac{m_c}{m_0} =\frac{V_0+\sqrt{V_0^2+ 2gz_{\Delta t}}+g\Delta t}{V_{\Delta t}} \ .
\label{eqn:addedmassratio}
\end{equation}

\noindent    We can calculate a lower bound on $m_c$  with some constraints.   The velocity change over the ballistic free-fall motion $V_0+\sqrt{V_0^2+ 2gz_{\Delta t}}$ can be replaced with $g\Delta t_f$, where $\Delta t_f$ is the time of flight of a person when they are moving ballistically  between steps.    In cases where the added mass effect is strong enough to slow the impactor,  it must be true that $V_a < V_I$ at the end of the impact.  Any persistent running gait also requires that the center-of-mass velocity $V_0 > 0$ at the time of impact, where an equality would correspond to an optimized running in which the foot barely clears the surface of the suspension when lifting off from depth $z_{\Delta t}$ and  landing at the surface again for the next step.  With these constraints, we can put a simple lower bound on the  required added mass 

\begin{equation} 
 \frac{m_c}{m_0} > \frac{g(\Delta t+\Delta t_f)}{V_I}\ .
\label{eqn:addedmassratioboundlow}
\end{equation}

 A lower bound on the required fluid height $H_c$ can be obtained by plugging in some numbers for a  typical running person;  $V_I \approx 2.0$ m/s  (see Fig.~\ref{fig:stoppingdistance}, or \cite{WSBW00}), a minimum step duration $\Delta t \approx 0.15$ s, and a time of flight between steps of $\Delta t_f = 0.15$ s when running \cite{WSBW00}.  This  gives a lower bound on  the required added mass $m_c/m_0 > 1.5$  for a person to run on the surface. Inputting this into Eq.~\ref{eqn:addedmass_bound} gives a lower bound on the required fluid height of $H_c > 0.6$ m  that a person can run across due to the added mass mechanism.    


In terms of human running parameters, we only needed to make an assumption about the step duration $\Delta t$ and the time of flight $\Delta t_f$, which are remarkably consistent from person to person, and only get slightly larger as running speeds decrease \cite{WSBW00}.  Other aspects  of human running, like bending or energy storage in the leg, gait, etc. -- while relevant to running -- cannot get around this bound.  Putting in realistic values for $V_0$, $z_{\Delta t}$, and $V_{\Delta t}$ into Eq.~\ref{eqn:addedmassratio} would raise this lower bound  on the minimum fluid height $H_c$.  On the other hand, the estimate for $H_c$ is fairly insensitive to the rather small range of reasonable parameters for human running since $H_c \propto m_c^{1/3}$ in Eq.~\ref{eqn:addedmass_bound} for $H \gg D/2$.  Thus we expect this lower bound to be a good first order estimate for $H_c$.

 \subsection{Using trajectories to obtain more precise conditions for running on cornstarch and water due to the added mass effect}
\label{sec:bounds_trajectories}

To obtain a more precise prediction for the conditions where the added mass effect could explain the ability to run on cornstarch and water, we consider the velocity trajectories from Eq.~\ref{eqn:nondimensional}  in addition to the constraint of Eq.~\ref{eqn:addedmassratio}.   Equation \ref{eqn:nondimensional} assumes that during the impact, the person  and the added mass have the same center of mass. Realistically, after impact the leg usually bends to shorten the distance between center of mass and foot, slowing down the foot and the added mass relative to the center of mass of the person.  Thus, our calculation overestimates $V_a$ and the force  from the added mass effect during the impact, and still provide lower bounds on the minimum fluid height $H_c$, as well as a lower bound on the minimum impact velocity $V_c$ required for a person to run on the surface of cornstarch and water due to the added mass effect.

 \begin{figure}  
\centering
{\includegraphics[width=0.475 \textwidth]{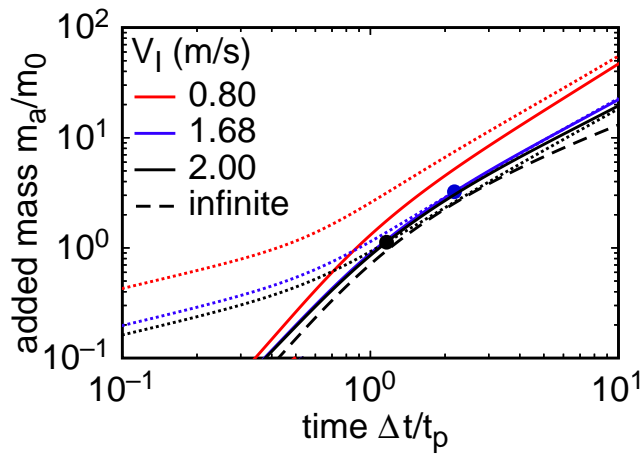}}
\caption {
(color online) Comparison between the produced added mass $m_a$ in response to impact and the required added mass $m_c$  for a person to run on the surface of cornstarch and water.  Curves are shown for $k=12$ and $V_0=0$. Solid line: produced added mass $m_a$ from Eq.~\ref{eqn:addedmass_dimensionless} for  impact velocities $V_I$ shown in the legend. Dashed line: produced added mass in the limit of $V_I\rightarrow \infty$.   Dotted lines: required added mass $m_c$, using the same color code for $V_I$ as the solid lines.  The solid points indicate the time $t_c/t_p$ where the produced added mass first exceeds the required added mass, thus meeting the threshold for a person to run on the surface.    
} 
\label{fig:addedmassreq}
\end{figure}

 To  numerically calculate a criteria for a person to run on  the surface of cornstarch and water, we compare  the  produced added mass over time from Eq.~\ref{eqn:addedmass} with the required added mass to support sustained running from Eq.~\ref{eqn:addedmassratio}.   To provide a constraint on the required added mass independent of the details of human running parameters,  we again assume $V_0 > 0$ in Eq.~\ref{eqn:addedmassratio} to obtain the lower bound 

 \begin{equation} 
 \frac{m_c}{m_0} > \frac{g \Delta t+\sqrt{2gz_{\Delta t}}}{V_{\Delta t}} 
\label{eqn:addedmassratiobound}
\end{equation}

\noindent  as a function of step duration $\Delta t$.  In contrast to the approximation in Eq.~\ref{eqn:addedmassratioboundlow}, this time we put the bound in terms of $z_{\Delta t}$ and $V_{\Delta t}$ which can be obtained from the trajectories from Eq.~\ref{eqn:nondimensional}.

\subsubsection{Critical velocity $V_c$}

At any given combination of $V_I$ and step duration $\Delta t$,   the added mass effect will provide enough  impulse to  allow persistent running if the produced added mass $m_a/m_0$  numerically integrated from Eqs.~\ref{eqn:nondimensional} and \ref{eqn:addedmass_dimensionless} exceeds the required added mass $m_c/m_0$ from Eq.~\ref{eqn:addedmassratiobound}.  
The numerical integration of Eqs.~\ref{eqn:nondimensional}  and \ref{eqn:addedmass_dimensionless} also produces the trajectories of $V_a$, and $z$ vs. $t$ that can be evaluated at time $\Delta t$ for Eq.~\ref{eqn:addedmassratiobound}.
These two expressions for the added mass are plotted as a function of step duration $\Delta t$ at $k=12$ in Fig.~\ref{fig:addedmassreq} for a few different $V_I$. The required added mass is shown as dotted lines, and the produced added mass is shown as solid lines.  There is a limiting curve for the produced added mass as $V_I\rightarrow \infty$.  The times $t_c$ where the produced added mass first exceeds the required added mass are  indicated by solid circles in Fig.~\ref{fig:addedmassreq}.    Once the produced added mass exceeds the required value for a given $V_I$ at time $t_c$, it tends to stay above that in the model calculation,  except for some slight variations around the threshold very near the threshold velocity. Thus, $t_c$ determines the minimum step duration required to satisfy Eq.~\ref{eqn:addedmassratiobound} and allow a person to run on the surface of cornstarch and water.

 \begin{figure}  
\centering
{\includegraphics[width=0.375 \textwidth]{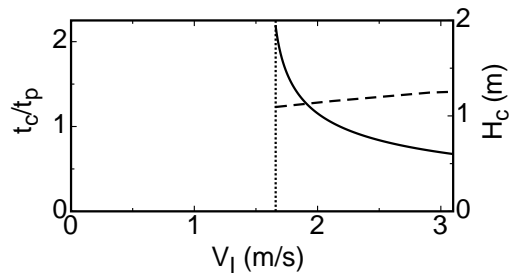}}
\caption {
The minimum step duration $t_c$ (solid line) required  for a person with foot impact velocity $V_I$  to run on the surface of cornstarch and water due to the added mass effect.   The divergence identifies a lower bound on the minimum foot impact velocity $V_c=1.66$ m/s,  shown as the  dotted vertical line.    The right-side scale shows the minimum fluid height $H_c$ (dashed line) required  for a person to run on the surface of cornstarch and water with a step duration $\Delta t=0.15$ s. 
} 
\label{fig:criticalvelocity}
\end{figure}

From the curves shown in Fig.~\ref{fig:addedmassreq}, it is apparent that at the smaller value of $V_I$, the produced added mass $m_a$ never exceeds the required added mass $m_c$.  A plot of  the  minimum time $t_c/t_p$  where the produced added mass $m_a$ first exceeds the required added mass $m_c$ is shown as a function of $V_I$ in Fig.~\ref{fig:criticalvelocity}.  We find that there is a critical velocity $V_c$ that is a lower bound on the minimum impact velocity required for $m_a > m_c$ and thus for a person to run on the surface of the suspension.   This appears as a critical point in the sense that $t_c/t_p$   diverges as $V_c$  is approached from above.     This critical  velocity is the threshold where the added mass effect is just strong enough to overcome gravity over the course of the impact.  The value of $V_c = 1.66$ m/s for $k=12$ (appropriate for the weight fraction range $0.57 \le \phi \le 0.61$ \cite{MMASB17}) happens to be between typical jogging and running values of $V_I$, which may be relevant for explaining why pools of cornstarch and water have been easily made in which people can run but not jog on the surface \cite{youtube_running}.

\subsubsection{Dependence on running gait}

 \begin{figure}  
\centering
{\includegraphics[width=0.375 \textwidth]{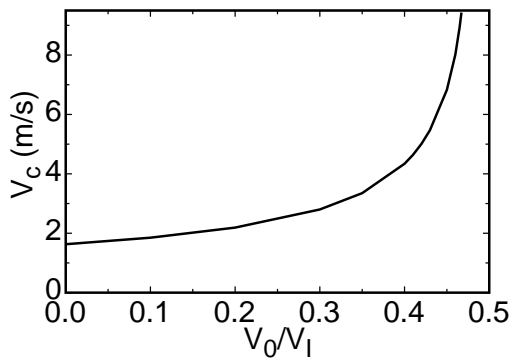}}
\caption {
Dependence of the critical foot impact velocity $V_c$ on running gait, represented by the ratio of center-of-mass velocity $V_0$ to foot impact velocity $V_I$.  A strong dependence on running gait is found, with a critical ratio $V_0/V_I=0.47$ where $V_c$ diverges.  A typical running gait on flat stiff ground corresponds to $V_0/V_I=0.28$ and $V_c=2.6$ m/s.
} 
\label{fig:VC_V0}
\end{figure}

The existence of the critical impact velocity $V_c$ as a bound is independent of any parameters of human running.  However, the value of the threshold velocity varies with the center-of-mass velocity $V_0$ at the time of impact, which depends on the running gait.  This can be  quantitiatively accounted for in terms of the ratio of  center-of-mass velocity  $V_0$ to foot impact velocity $V_I$. When $V_0/V_I=0$, this corresponds to an optimal gait for running on cornstarch and water, corresponding to an equality in Eq.~\ref{eqn:addedmassratiobound}.  In the other limit, $V_0/V_I=1$ corresponds to a rigid leg just before impact.  The critical velocity $V_c$  is shown as a function of $V_0/V_I$  for $k=12$ in Fig.~\ref{fig:VC_V0}, where $m_c$ is calculated with positive (downward) values of $V_0$ from Eq.~\ref{eqn:addedmassratio} instead of Eq.~\ref{eqn:addedmassratiobound}. The threshold velocity $V_c$ increases with $V_0/V_I$. At $V_0/V_I=0.47$, there is a critical point where the required impact velocity $V_c$ diverges.  Just as in Fig.~\ref{fig:criticalvelocity}, this is a threshold where the added mass effect is just strong enough to overcome gravity over the course of the step, and here must additionally overcome  the initial momentum of the person moving down at center-of-mass velocity $V_0$.  This means the specific running gait can have a significant effect on whether a person can run on the surface of cornstarch and water.   For a  rigid leg before impact where $V_0=V_I$, running on the surface of cornstarch  and water  would not be possible for any impact velocity $V_I$ due to this critical point.    For typical running on flat, stiff ground, ballistic motion between steps results in $V_0=g\Delta t_f/2 =0.75$ m/s for a time of flight $\Delta t_f=0.15$ s.  The critical foot impact velocity for this value of $V_0$ would be $V_c =2.6$ m/s corresponding to $V_0/V_c$=0.28. While this is a fast foot impact velocity for a runner, it is in an achievable range, especially  if a person moves their legs faster than they would normally run on solid ground to increase $V_I$ relative to $V_0$.  

As a separate aspect of the gait, the trajectories shown in Fig.~\ref{fig:integration} assume the person keeps their leg rigid after impact to maximize the growth of the added mass region.  This is unnatural, as legs are usually bent in response to the impact of landing to minimize the stress on the body.  If a person runs with a more typical running gait where the legs bend in response to impact, the critical velocity $V_c$ would increase further beyond the values given above for more ideal gaits.  This is not modeled quantitatively here because the response depends on the mechanics of the person, which is beyond the scope of this paper.

 \subsubsection{Step duration}
 
In contrast to objects in free fall, a running person lifts their foot, so after the step duration $\Delta t$,  the added mass effect no longer applies to that foot, and so the trajectories in Fig.~\ref{fig:integration} no longer apply at times later than $\Delta t$.  Thus, for a person to run on the surface of cornstarch and water, the step duration $\Delta t$ must be larger than the time $t_c$ required for the produced added mass to exceed the required added mass identified in Figs.~\ref{fig:addedmassreq} and \ref{fig:criticalvelocity}.    We find a typical scale for $t_c/t_p \approx 1$ when $V_0=0$ for all $V_I>V_c$ except very near  $V_c$ (Fig.~\ref{fig:criticalvelocity}).  We also find $\Delta t \gg t_p$ over the parameter range studied. Thus, $\Delta t \gg t_c$, and most of the deceleration in the trajectories of Fig.~\ref{fig:integration} applies to the person.  This means the  minimum velocity condition  for running on the surface is met as long as $V_I$ is even just a little bit above $V_c$.  A correction to $V_c$ accounting for the value of $\Delta t$ is less than 1\% for  $\Delta t =0.15$ s, $V_I =2$ m/s and $k=12$.    For 
$V_0 > 0$, $t_c/t_p$ increases, and $t_c$ becomes greater than the typical step duration $\Delta t=0.15$ s for $V_0>0.41V_I$.  This step duration constraint may never be relevant for a person running on cornstarch and water, however, because at these parameters the critical foot impact velocity is $V_c=4.6$ m/s, a much harder constraint to meet.

\subsubsection{Minimum fluid height $H_c$}

The minimum fluid height $H_c$ required for a person to run on the surface of cornstarch and water calculated in Sec.~\ref{sec:bounds} can be updated to account for the velocity trajectories. $H_c$ is calculated from Eq.~\ref{eqn:addedmass_bound} for the required added mass $m_c$ from Eq.~\ref{eqn:addedmassratiobound}, using the trajectories of Eq.~\ref{eqn:nondimensional} to obtain $z_{\Delta t}$ and $V_{\Delta t}$. This is shown  on the right-side scale of Fig.~\ref{fig:criticalvelocity} for a typical step duration of $\Delta t=0.15$ s.  This yields $H_c=1.2$ m at $V_I=V_c$.  The critical height $H_c$ is seen to be relatively insensitive to $V_I$.   This value of $H_c=1.2$ m is twice that calculated from the simpler bound of Eq.~\ref{eqn:addedmassratioboundlow}.
 
The large typical ratio of $\Delta t/t_c$ for human running parameters has a significant effect on $H_c$, since the added mass keeps growing after the minimum time $t_c$ required to run on the surface of cornstarch and water, and if the dynamically jammed region collides with a boundary during this time, the added mass effect dies out.  This means that for the added model to describe the ability of a person to run on the surface of cornstarch and water, significantly more added mass  is required  than if the impact duration $\Delta t$ was tuned (i.e.~for an optimized machine) to its minimal value $t_c$.  For example, for a runner with $\Delta t = 0.15$ s, $V_I=2$ m/s, $V_0=0$ at  $k=12$, then $\Delta t=6.5t_p$; at these conditions,   $m_a = 12m_0$,  and the required added mass according to Eq.~\ref{eqn:addedmassratiobound} is $11m_0$,  only slightly below the actually achieved added mass.   This requires a pool of minimum height $H_c=1.2$ m according to Eq.~\ref{eqn:addedmass_bound}.  However, if $\Delta t$ was tuned  to be near $t_c$, the required added mass would be only $1.2m_0$ as seen in Fig.~\ref{fig:addedmassreq}, corresponding to $H_c=0.6$ m.

\subsubsection{Penetration depth}

 For a pool of height $H> H_c$, the  maximum foot penetration depth $z_{\Delta t}$  can be calculated from Eq.~\ref{eqn:nondimensional} at the time of foot liftoff $\Delta t$.   Based on the typical scales observed in Fig.~\ref{fig:integration}, the typical length scale of the penetration depth scales $z_0/k$.   More precisely, for a typical running step duration of $\Delta t \approx  0.15$ s, $V_I=2.0$ m/s, and an ideal gait with $V_0=0$ at $k=12$, the penetration depth is $z_{\Delta t} = 2.1z_0/k = 10$ cm.  This is a step height that is reasonable to achieve while running.  It also corresponds to a minimum time of flight  $\Delta t_f=\sqrt{2z_{\Delta t}/g} = 0.12$ s for $V_0=0$, near typical times for runners \cite{WSBW00}, so that it could be achieved without need for a large $V_0$.  $z_{\Delta t}$ is  insensitive to most parameters. For example, doubling $V_I$ only increases $z_{\Delta t}$ by $17\%$.  Thus, the penetration depth is not likely to be a limiting constraint on the ability to run on the surface of cornstarch and water.
 
For the experiments shown in Fig.~\ref{fig:stoppingdistance}, the added mass effect is much more limited since $H \ll H_c$.  The energy balance calculation in Eq.~\ref{eqn:energybalance} can be modified by adding the work $W_a$ done by added mass effect.  The force from the added mass effect can be expressed as the time-derivative of the momentum to obtain the work $W_a = \int [d(m_a V_a)/dt] dz$.  Modifying Eq.~\ref{eqn:zs} accordingly leads to a correction to the penetration depth $z_s$  by no more than 2\% for the parameters in Fig.~\ref{fig:stoppingdistance}.  Not only is the added mass effect not enough to account for running on cornstarch and water for small $H$, but it also is a negligible compared to the contribution of the system-spanning dynamically jammed region for small $H$.

\subsubsection{Weight fraction dependence}

 \begin{figure}  
\centering
{\includegraphics[width=0.475 \textwidth]{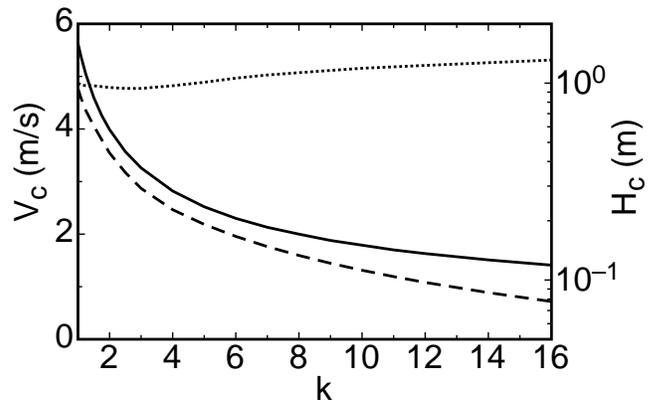}}
\caption{Solid line: the critical foot impact velocity $V_c$ vs. $k$, where  larger $k$ corresponds to larger weight fraction $\phi$.  A typical running foot impact velocity of $V_I=2$ m/s sets a lower limit of $k=8$ ($\phi=0.56$) at which a typical person can run on the surface of cornstarch and water.  The right-hand scale shows the maximum foot penetration depth $z_{\Delta t}$ (dashed line) and the  minimum fluid height $H_c$ (dotted line), both at $V_I=2$ m/s.  
} 
\label{fig:phidependence}
\end{figure}

To describe trends in weight fraction $\phi$, we use the velocity ratio $k$ as  a proxy, since it is the material parameter in the model that depends on $\phi$.  $k(\phi)$ was found to increase with $\phi$ up to a plateau of $k=12\pm 4$ at high weight fractions ($0.57 < \phi < 0.61$), where the liquid-solid transition occurs at $\phi_c=0.61$  \cite{MMASB17}. We plot values of the critical impact velocity $V_c$ vs.~$k$ in Fig.~\ref{fig:phidependence}, assuming an ideal gait with $V_0=0$. This calculation is probably not realistic for small $k$ close to 1, where the existence of a larger plug-like flow in front of the impactor \cite{ASMMB17} may have a significant effect on the expression for the growth of the added mass over time. At larger $k$ where the model does apply, a minimum weight fraction for which people can run on the surface of cornstarch and water can be obtained as the $k$-value where $V_c =2$ m/s  as  a typical foot impact velocity for a running person.  We find  $V_c=2$ m/s when $k=8$,  yielding a minimum  weight fraction of $\phi=0.56\pm 0.01$ \cite{MMASB17}. Because  of the steep variation of $k(\phi)$ in this range \cite{MMASB17}, this 0.01 error includes the large run-to-run variation of 30\% in $k$.  In the high-weight-fraction plateau where $k=12\pm 4$,  $V_c$ only varies as $V_c=1.7\pm 0.3$ in this range due to the relative insensitivity of $V_c$ to $k$ in Fig.~\ref{fig:phidependence}.  This identifies a robust minimum velocity at high weight fractions, remaining in between typical jogging and running velocities.  

The maximum penetration depth $z_{\Delta t}$ also changes with weight fraction.  This is plotted on the right-side scale of Fig.~\ref{fig:phidependence} at the threshold $V_I=V_c$, $V_0=0$, and $\Delta t=0.15$ s.  In the $k$-range where $V_c < 2$ m/s,   this gets as large as $z_{\Delta t}=0.13$ m at $k=8$, which is a reasonable height for a person to step, so $z_{\Delta t}$ never becomes the limiting factor in running on cornstarch and water.  At smaller $k$, where $V_c$ becomes larger than human foot velocities, $z_{\Delta t}$ also reaches prohibitively larger values.  

Finally, we plot the required fluid height $H_c$ vs.~$k$ on the right-side scale of Fig.~\ref{fig:phidependence} for the same threshold $V_I=V_c$, $V_0=0$,  and $\Delta t=0.15$ s. $H_c$ is also insensitive to $k$, and varies only as $H_c=1.2 \pm 0.1$  for $k=12\pm 4$.

\section{Conclusions}

We tested a constitutive model for the impact response of suspensions consisting of an effective stiffness per unit area of the system-spanning dynamically jammed region after a delay time required for it to span the system \cite{MMASB17}.  We tested this model by applying it to a person walking and running on the surface of cornstarch and water, in which the velocity varies, and the impactor shape is different from the experiments the constitutive model was fit to.  The model predicts the penetration depth of free-falling objects into suspensions, in which the kinetic energy of the impacting object is balanced by the work required to deform the system-spanning dynamically jammed region.  The model prediction of the penetration depth of a person's foot walking, jogging and running on the surface of cornstarch and water suspensions is within the run-to-run standard deviation of stress measurements of 30\%. In addition, the model explains an observed delay before the foot comes to a sharp stop, arising from the time it takes for the front of the dynamically jammed region to reach the boundary opposite the impactor.  This feature could not be explained by models with a smoothly varying stress response such as a bulk viscosity or the added mass effect.

We  performed detailed calculations of the added mass  contribution to impact response based on the model of Waitukaitus \& Jaeger \cite{WJ12}.  We showed that it could be strong enough to explain the ability of a person to run on the surface of cornstarch and water, but only if the fluid is deeper than $H_c=1.2\pm0.1$ m and the impact velocity is greater than $V_c = 1.7\pm0.3$ m/s. These bounds apply where the effect is strongest in the weight fraction range $0.92\phi_c \le \phi \le \phi_c$.  These bounds assume  an optimal gait where the center-of-mass velocity of the person $V_0=0$  at the time of impact.  $V_c$ increases for either smaller $\phi$ or larger $V_0$, although $H_c$ is fairly insensitive to any parameters of human running.  

 It is notable that the critical velocity $V_c=1.7$ m/s  for a wide range of weight fraction is in between typical impact velocities for jogging and running.  This could explain why there are many examples of pools of cornstarch and water that people can run but not jog on the surface.   On the other hand, this may be a coincidence,  as it is relatively easy to tune the desired thickness of the suspension by hand, and the system-spanning dynamically jammed region can also  explain the ability to run on cornstarch and water in this parameter range.  Since most videos readily available  have been taken without controlled laboratory conditions \cite{youtube_running},  it is not always clear if $H>H_c$, and  more detailed measurements  are not available to test the two models in such cases, especially since they both predict comparable penetration depths, and can predict packing-fraction-dependent minimum velocities.  The one clear-cut set of experiments we have analyzed in Fig.~\ref{fig:stoppingdistance} and shown in Supplementary Video 1 clearly shows  people walking, jogging, and running on cornstarch at $H=0.1$ m ($< H_c$) and foot  impact velocities as low as $V_I =0.2$ m/s $(< V_c)$.   In  this range of $H<1.2$ m and $V_I<1.7$ m/s, only the system-spanning dynamically jammed region is able to explain  why people can run and walk on cornstarch and water.   Measurements of people running on  pools of fluid deeper than $H_c=1.2$ m could confirm whether the added mass mechanism can  in some cases explain the ability of people to run on the surface of cornstarch and water.

 The fact that we were able to use an averaged constitutive model based on constant velocity impacts to successfully predict the mechanics of the suspension when a person runs on the surface indicates that the constitutive model can be used on flows with different geometry and driving conditions than were originally used to develop the model.  This  opens up the possibility that the  constitutive model may be applicable to a wider  variety of problems with different forcing conditions, boundary conditions, and flow geometries.  This is a non-trivial result, as the traditional method of obtaining constitutive relations based on steady-state rheometer measurements has not yet been able to explain any transient or dynamic phenomena of DST suspensions besides steady shear, as least in the absence of time-dependent hysteresis terms \cite{BJ14}.  One such phenomenon is the oscillation of the velocity of a sphere sinking in a  suspension, rather than monotonically approaching a terminal velocity as it would in a generalized Newtonian fluid \cite{KSLM11, KSM13, GBMP17}. It was argued that a repeated process of jamming and unjamming of something like  the dynamically jammed region could account for such oscillations  \cite{KSLM11, KSM13, GBMP17}.  Now we have an averaged constitutive model that includes such a process, along with a relaxation process to describe the unjamming \cite{MB17}.  Similarly, it was shown that the formation of stable holes in the surface of a vertically vibrated layer of a DST suspension could not be explained by a steady-state rheology in the absence of hysteresis in the constitutive relation $\tau(\dot\gamma)$ \cite{De10}. This apparent hysteresis appears to be time-dependent when the constitutive relation is put in terms of $\tau(\dot\gamma)$.  Alternatively, such hysteresis could come about for a history of increasing shear rate from the delay time we observed before the large stress increase.  For the history of decreasing shear rate, the the time dependence may come from the relaxation time of the dynamically jammed state \cite{MB17}.  Finally, the observation of objects bouncing off the surface of a DST suspension remains unexplained based on steady-state or added mass models  which are dissipative constitutive relations \cite{BJ14, WJ12}.  The system-spanning dynamically jammed region  can  in principle provide some energy storage in the modulus $E$ that could possibly explain the ability of impacting objects to bounce off the surface.   Testing the general applicability of the constitutive relation to  these and other problems is left open for future work.

\section{Acknowledgements}

We thank Scott Waitukaitis, Ivo Peters, Heinrich Jaeger and Madhusudhan Venkadesan for discussions and for sharing their unpublished results. This work was supported by the NSF through grant DMR 1410157.

\section{Supplementary Videos}

Supplementary videos may be downloaded from:
\newline https://www.eng.yale.edu/brown/publications.html

\end{document}